\documentstyle[prl,multicol,aps]{revtex}

\begin{document} 
\draft
\twocolumn
\title{Simplified dynamics for glass models.}
\author{D.B. Saakian}
\address{Yerevan Physics Institute,
Alikhanian Brothers St. 2, Yerevan 375036, Armenia \\ Universidad de La Frontera, Departamento de Cienas Fisicas,  Casilla
54-D, Temuco, Chile.}

\maketitle
\begin{abstract}
In spin glass models one can remove minimization of free energy by some order
parameter. One can consider hierarchy of order parameters. It is possible to
divide energy among these parts. We can consider relaxation process in glass
system phenomonologically, as exchange of energy between  2 parts. It is
 possible to identify trap points in phase space.
We suggest some phenomonological approximation-truncated Langevine. 
 The mean field statics is used to introduce a
 phenomenologic dynamics as its natural extension. 
 Purely kinetical phase transitions are investigated..

\end{abstract}                   
\vspace{10mm}

One of the most striking features of nature is hierachy. They are a lot-
elementary particles, nucleai, atoms, kinetics, hydrodynamics...
Such hierchial structure is common for complex systems. These hierachial levels are almost independent. It is interesting to consider a simple toy model and investigate interconnection of thermodynamics at subsequent levels.
Such objects are glasses and spin glasses [1,2]. Here  
the hierarchy of degrees of freedom and even hierarchy of temperatures  arise naturally [3].
Let us consider the model of $p$-spin spherical glass [4], when couplings are also dynamical with different temperature. This has been considered in [5]
for the partial anealing phenomena [6]. What we modify, couplings do not have  any own hamiltonian. The gaussian integration for them we consider as an integration measure only (otherwise they could have to much entropy, noncomparable with entropy of spin
s):
\begin {equation}
\label{b1}
D(J)=
\sqrt{\frac{2M}{N\pi}}
\exp \left ( - \sum_{i_1..i_p}    \frac{2M}{N}    j^2_{i_1..i_p} \right )  \prod_{i_1..i_p}{\it
d}j_{i_1..i_p}
\end{equation}
Here $M\equiv C^p_N=N!/p!(N-p)!$ is a total number of couplings.
We can define then some partition
\begin {equation}
\label{b2}
Z=\int
D(J)\left  [   {\rm Tr}   _{s_i}   \exp \left (-\sum_{i_1..i_p}j_{i_1..i_p}
s_{i_1}..s_{i_p}  \right )   \right  ]^n
\end{equation}
According to [5], 
\begin {equation}
\label{b3}
n=\frac{\beta_j}{\beta}
\end{equation}
where inverse temperature $\beta_j,\beta$ correspond to spins and couplings.
Then it is possible to derive [7]
\begin {equation}
\label{b4}
-E_t=\frac{1}{n}  \frac{  \partial      }{    \partial    \beta} \ln Z(n,  \beta), \  \
-E_j= \frac{  \partial      }{      \partial   n}T\ln Z(n, \beta)
\end{equation}
Here $E_t$ is a total energy, $E_j$ is connected with the slow motion of couplings. We assume, that spins 
are attached to thermostat with one temperature, couplings-to other.
One usually  considers heat exchange between seperate objects. In our case we seperate slow and
fast variables and our goal should be calculation of heat current between them. We consider such
simplified version of relaxation process: phenomonological heat exchange between fast and slow variables.
 Even on such level without any specification of dynamics one can found something important for the dynamics.
 When we isolate our system from thermostats, we ascribe to our system at  any moment of time 2 temperatures
 (main assumption).
So our system evolves along  the line 
 \begin {equation}
\label{b5}
E_t=const,\quad d E_t(\beta,n)\equiv \frac{d E_t}{d \beta}d \beta+\frac{d E_t}{d n} d n=0
\end{equation}
If there is a point, where we have also
 \begin {equation}
\label{b6}
d E_j(\beta,n)\equiv \frac{d E_j}{d \beta}d \beta+\frac{d E_j}{d n} d n=0
\end{equation}
then system could get trapped in such point (glassy phenomena). For such points we have an equation
 \begin {equation}
\label{b7}
\frac{d E_t}{d \beta}\frac{d E_j}{d n}-\frac{d E_j}{d \beta}\frac{d E_n}{d n}=0
\end{equation}
Let us now specify some simplified dynamics. The main point is to define heat currenet between spins and couplings at given temperatures.
Having its expression we can calculate everything assuming a picture, where couplings interact only with spins, while spins could interact
with external fields or thermostats.
Let us postulate some simplified dynamics with a single parameter (eefective temperature $\beta_0$).
Actually we see, that dynamic equations for any thermodynamic variable is 
maximal simple.
 For any function $g(\{j_k\})$ and its average G,
\begin{equation}
\label{b8}
G(\beta,n)=\langle g(\beta,n)\rangle
\end{equation}
 we can write an equation:
\begin{equation}
\label{b9}
\frac{d G}{dt
}=\langle \sum _k (g''_{
j_kj_k}(n,\beta,j)+
g'_{ j_k}(n,\beta,j) f'_{j_k}(1,\beta_0,j))         \rangle
\end{equation}
where the average is taken by the distribution $D(j) Z^n (j, \beta) $
Here $f(n,\beta)\equiv Z(\beta,j)$ is a free energy of our partial anealing model. 
Let us put a constraint, that total energy is not changed.
\begin{equation}
\label{b10}
-\frac{d E_t}{d t}=  \langle       \sum _k  (f'''_{\beta
j_kj_k}(n,\beta,j)+
f''_{\beta j_k}(n,\beta,j) f'_{j_k}(1,\beta_0,j))     \rangle
\end{equation}
This is a simplest differential equation, that we can write. The wright part 
disappears,at $n=1$, when $\beta=\beta_0$. So the choice of first term in (10) defines the form of second term. Solving this equation we can find this effective temperature $\beta_0$ as a function of $(n,\beta)$, then calculate evolution of $E_j$:
\begin{equation}
\label{b11}
-\frac{d}{dt} E_j=T\langle     \sum _k  (f'''_{n
j_k  j_k}   (   n,\beta ,j  )  +
f''_{n j_k}    (n,   \beta  ,   j    ) f'_{j_k}   (1,\beta_0,j))   \rangle
\end{equation}
Here both terms are the same order. 

For the free energy of $p=2$ model  we have  replica symmetry solution [4]
\begin{eqnarray}
\label{b12}
&& f=n\frac{1+\ln 2\pi}{2}+\frac{\beta^2}{4}[(n^2-n)q^2+n]\nonumber \\
&&+\frac{n\ln(1-q)}{2}+\frac{1}{2}\ln[1+\frac{nq}{1-q}]
\end{eqnarray}
 There is one solution $q_1=0$. The equation for two other solutions: 
\begin{equation}
\label{b13}
(1-n)q^2+(n-2)q+1-1/\beta^2=0
\end{equation}
\begin{equation}
\label{b14}
q_{2,3}=\frac{(2-n)\pm\sqrt{(n-2)^2-4(1-n)(1-\beta^2)}}{2(1-n)}
\end{equation}
At low temperatures our system is in the PM phase with $q=0$. While frozen it could be transformed into SG phase. 
For the case $n<2$ the transition point is $\beta_c=1$ and we take the solution $q=q_2$. For $n>2$ 
the critical value of $\beta$ is changed. Having the values of $q$ we can define 
\begin{eqnarray}
\label{b15}
&&\frac{dq}{d\beta}=-\frac{2}{\beta^3[(1-n)2q-2+n]}, \frac{dq}{d n}=\frac{q^2-q}{[(1-n)2q-2+n]}\nonumber \\
&&\frac{d f}{d \beta}=\frac{\beta}{2}[(n^2-n)q^2+n]+[(n^2-n)/2\beta^2 q 
+\frac{(1-n)}{2(1-q)}\nonumber \\
&&-\frac{1-n}{2(1-(1-n)q})]q'_\beta \nonumber \\
&&\frac{d^2 f}{d n d \beta}=\frac{\beta}{2}[(2n-1)q^2+1]
+[(n^2-n)/2\beta^2 +\frac{(1-n)}{2(1-q)^2} \nonumber \\
&&-\frac{(1-n)^2}{2(1-(1-n)q)^2}]q'_\beta q'_n +[(2n-1)/2\beta^2 pq-\frac{1}{2(1-q)}\nonumber \\ 
&&+\frac{1}{2(1-(1-n)q)^2}]q'_\beta
+\beta(n^2-n)qq'_n   \nonumber \\
&&\frac{d^2 f}{d \beta^2}=\frac{1}{2}[(n^2-n)q^p+n]+
[(n^2-n)/2\beta^2 +\frac{(1-n)}{2(1-q)^2}\nonumber \\
&&-\frac{(1-n)^2}{2(1-(1-n)q)^2}](q'_\beta)^2  
+2\beta(n^2-n)qq'_\beta
\end{eqnarray}
We have for energies
\begin {equation}
\label{b16}
-E_t(q,n,\beta)=\frac{\beta}{2}[(n-1)q^2+1]
\end{equation}
$$-E_j(q,n,\beta)=\frac{\ln(1+\ln 2\pi)}{2\beta}$$
$$+[(2n-1)q^2+1]\frac{\beta}{4}+\frac{\ln(1-q)}{2\beta}
+\frac{q}{2\beta(1-q+nq)}$$

The trap point could be defined nullifying the determinant of the linear system:
\begin {equation}
\label{b17}
[(n-1)q^2+1]d\beta+\beta q^2dn+\beta(n-1)2qdq=0
\end{equation}
$$\{-[\ln(1+\ln 2\pi)
+\frac{q}{(1-q+nq)}+\ln(1-q)]\frac{1}{\beta^2}$$
$$+\frac{[(2n-1)q^2+1]}{2}\}d \beta+[\beta q^2-\frac{q^2}{\beta(1-q+nq)^2}]d n$$
$$+\{(2n-1)\beta q-\frac{1}{\beta(1-q)}
+\frac{1}{(1-q+nq)^2\beta}\}d q=0$$
$$2d\beta/\beta^3-(q^2-q)dn+[2q(1-n)+n-2]dq=0$$

To calculate (11) we need to consider the model with $m+n$ replicas, when $m\to 0$. 
 in $s=<s_{\alpha}s_{\beta}>,1\le \alpha\le n,1\le m, m\to 0$ and $r=<s_{\gamma}s_{\beta}>,1\le \gamma\le m$
 We  derive 
\begin{equation}
\label{b18}
\beta (1-q)=\beta_0(1-r)
\end{equation}
\begin{equation}
\label{b19}
s^2=q[1-\frac{\beta(1-q)}{\beta_0}]
\end{equation}
\begin{equation}
\label{b20}
\frac{d s}{d n}=\frac{1}{2s}[1-\frac{\beta}{\beta_0}+\frac{2\beta q}{\beta_0}]\frac{d q}{d n}
\end{equation}
$$\frac{d s}{d \beta}=\frac{1}{2s}[1-\frac{\beta}{\beta_0}+\frac{2\beta q}{\beta_0}]\frac{d q}{d \beta}-\frac{q(1-q)}{\beta_0}$$
The set of equations (24-27) describe one mode of solution. For the small values of $\beta_0$, when $\beta_0<\beta(1-q)$ works another solution
\begin{equation}
\label{b21}
r=0,
s=0
\end{equation}

Let us consider the dynamics. 
We should consider the dynamics with the effective hamiltonian
\begin {equation}
\label{b22}
h+f\equiv-\sum_{i_1..i_p}2M/(N)j^2_{i_1..i_p}+f(j,\beta)
\end{equation}
using the (9),(10) we derive 
To be short we will use 
$$j_k=j_{i_1..i_p}$$
\begin {equation}
\label{b23}
-dE_t/dt=\sum_kf'''_{\beta j_kj_k}(j,\beta)+f''_{\beta j_k}(j,\beta)(h'_{j_k}+f'_{j_k}(\beta_0))
\end{equation}
\begin {equation}
\label{b24}
-\beta dE_j/dt=1/\beta\{\sum_kf'''_{n j_kj_k}(j,\beta)
\end{equation}
$$+f''_{n j_k}(j,\beta)(h'_{j_k}+f'_{j_k}(\beta_0))\}$$
To be short we will use  $j_k=j_{i_1..i_p}$
\begin {equation}
\label{b25}
- dE_t/dt=[2\beta(1-q^2)-2\beta^2q^{p}\frac{d q}{d \beta}+\beta_0s^p+
\end{equation}
$$2\beta\beta_0 s^{p}\frac{d s}{d \beta}]M
-4M/N[\beta f''_{\beta\beta}+f'_\beta]=0$$
Then we can caculate heat current
\begin {equation}
\label{b26}
-\beta dE_j/dt=[-\beta^2pq^{p-1}\frac{d q}{d n}+
\end{equation}
$$\beta\beta_0 ps^{p-1}\frac{d s}{d n}]M\\
-4M/N\beta f''_{\beta n}$$
Having this expression one can define different relaxation processes.\\
So in our system we have three temperatures- two for statics, one for dynamics. It is very interesting the situation, when the statics corresponds to SG phase, the dynamics- to PM. A rich physics is possible.\\
It is possible to construct similiar dynamics for SG models in external magnetic field. Here the role of $J_k$ can play partial populations with different magnetization. Instead of $\beta_j$ one can consider the total magnetization M. It takes its equilib
rium value only after long relaxation process. Morever, one can remove the saddle point condition by some order parameter and consider it as free parameter. One can consider free energy as an energy for such (nonthermalized) order parameter. 
Again it is possible to construct effective energy for its relaxation dynamics and some effective simplified dynamics.
 So mean field approach allows besides statics to introduce some 
 phenomonologic dynamics as a natural extension.
If we have for a partition a representation
\begin{equation}
\label{a27}
Z=\int {\it d} m Z(m,\beta)=\int {\it d }m\exp{\ln Z(m,\beta)}
\end{equation}
We can connect with order parameter m  an effective hamiltonian (and energy)
\begin{equation}
\label{a28}
-\ln Z(m,\beta)/\beta
\end{equation}
 For a total energy of system
we have
\begin{equation}
\label{a29}
-E_t(m,\beta)=\frac{d}{d \beta}\ln Z(m,\beta)
\end{equation}
Again it is possible to construct some effective simplified dynamics.
 It is possible to repeat 
this procedure of thermalization several times, assuming the hierarchy of relaxation periods. \\
 So mean field statics allows besides statics to introduce some 
 phenomenologic dynamics as a natural extension. There is a nice chance of phase transition in kinetics. It will be interesting to consider protein folding under such point of view.
Another possible generalization-application in collision integrals of kinetic theory. Really we wrote (9), assuming only the smootness of distribution.
\begin{center} {\Large \bf Acknowledgements} \end{center}
I am grateful to
Fundacion Andes grant c-13413/1 for financial support,A.E. Allahverdyan,A.Crisanti, Th. Niewenhuizen. I thank L. Peliti  
for warm hospitality in Napoly and very fruitful discussions.


\begin{thebibliography}{99}
\bibitem{ni} M. Mezard,G. Parisi,M.A. Virasoro,Spin-glass Theory and Beyond,World Scientific,Singapore,1986.
\bibitem{ru}  K.Binder,A.P. Young, R.M.P. {\bf 58} (1986),801
\bibitem{ru}  Th. Nieuwenhuisen Phys. Rev. Let. {\bf 79} (1997),1317
\bibitem{ch} A. Crisanti,H.J. Sommers, Z. Phys. {\bf 87B},341
\bibitem{s1}  V. Dotsenko,M. Mezard,S. Franz J. Phys. A{\bf 27} (1994) 2351
\bibitem{so}  A.C.C. Coolen,R.W. Penney,D. Sherrington J. Phys. A{\bf 26} (1993) 3681
\bibitem{ar}  A.E. Allahverdyan,T.Nieuwenhuizen,D.B. Saakian (1999) 
\end{thebibliography}
\end{document}